\documentclass[twocolumn,showpacs,prl]{revtex4}
\newcommand{\ket}[1]{|#1\rangle} 
\newcommand{\bra}[1]{\langle #1|} 
\begin{document}
\title{Weak cloning of an unknown quantum state}
\author{Erik Sj\"oqvist}
\email{erik.sjoqvist@kvac.uu.se}
\author{Johan {\AA}berg}
\email{johan.aaberg@kvac.uu.se}
\affiliation{Department of Quantum Chemistry,
Uppsala University, Box 518, SE-751 20 Uppsala, Sweden}
\date{\today}
\begin{abstract}
The impossibility to clone an unknown quantum state is a powerful
principle to understand the nature of quantum mechanics, especially
within the context of quantum computing and quantum information. This
principle has been generalized to quantitative statements as to what
extent imperfect cloning is possible. We delineate an aspect of the 
border between the possible and the impossible concerning quantum
cloning, by putting forward an entanglement-assisted scheme for
simulating perfect cloning in the context of weak measurements. This
phenomenon we call weak cloning of an unknown quantum state. 
\end{abstract}
\maketitle
Although the no-cloning principle \cite{wootters82,dieks82} forbids
perfect cloning of an unknown quantum state, approximate cloning 
whose quality is independent of the input state is known to be
possible \cite{buzek96,linares02}. The prize which is paid in such 
approximate cloning procedures is that no state is cloned perfectly; 
all states are cloned by a maximum fidelity less than one \cite{bruss98}. 
In contrast, we here demonstrate a scheme that we shall call `weak
cloning', which allows independent local weak measurements 
\cite{aharonov90} as if performed on a perfectly cloned unknown 
input state.

Weak measurements on post-selected quantum ensembles leads to the
concept of weak values \cite{aharonov90}. Although conceptually
similar to expectation values, these weak values may show curious
behavior. For instance, they can lie far outside the range of allowed
values of the measured observables \cite{aharonov87,aharonov88} or
they can lie in classically forbidden regions of space
\cite{aharonov93a,aharonov93b}. Weak cloning is a phenomenon that
arises from weak measurements by post-selection.

To describe the idea of weak cloning, let us assume that two
observers, Alice and Bob, can do experiments on qubits. Alice is 
given a qubit prepared in the state $\ket{\phi}_1 = \alpha \ket{0}_1 + 
\beta \ket{1}_1$, where $\alpha$ and $\beta$ are unknown complex 
numbers with $|\alpha|^2 + |\beta|^2 = 1$. Alice shares also a 
Bell state, such as $\ket{\varphi_+}_{23} = \big( \ket{0}_2 \ket{0}_3 + 
\ket{1}_2 \ket{1}_3\big) /\sqrt{2}$, with Bob. Thus, initially the 
total state reads 
\begin{eqnarray} 
\ket{\Phi} = \ket{\phi}_1 \ket{\varphi_+}_{23} .   
\end{eqnarray}  
Alice and Bob perform weak measurements on qubits $1$ and $3$,
respectively. These measurements can be modeled by the interaction
Hamiltonian
\begin{eqnarray} 
H(t) & = & g_a(t) A \otimes I_2 \otimes q_a \otimes I_3 \otimes I_b
\nonumber \\ 
 & & + g_b(t) I_1 \otimes I_2 \otimes I_a \otimes B \otimes q_b ,
\label{eq:hamiltonian}  
\end{eqnarray}
where $q_a$ and $q_b$ are Alice's and Bob's pointer position variable,
respectively, $A$ and $B$ are the corresponding measured observables. 
The identity operators $I_1, I_2, I_a$ pertain to Alice's two qubits
and her pointer system, and $I_3, I_b$ are the identity operators of Bob's
qubit and pointer system. Note that the observable $A$ belongs to the
first of Alice's qubits, i.e., the one prepared in the unknown state
$\ket{\phi}$. The Hamiltonian in Eq. (\ref{eq:hamiltonian}) results
in an evolution that is local with respect to Alice's and Bob's
locations, and leaves Alice's second qubit unaffected. The
time-dependent coupling parameters $g_a$ and $g_b$ turn on and off the
interaction between the measurement devices and the measured
qubits. With the measuring devices initially in the product state
$\ket{m_a} \ket{m_b}$, we obtain the final total state as ($\hbar =1$)
\begin{eqnarray} 
\ket{\Gamma} & = & e^{-i\gamma_a A \, \otimes \, q_a} \otimes I_2  
\otimes e^{-i\gamma_b B \, \otimes \, q_b} \ket{\Phi} \ket{m_a} \ket{m_b}      
\end{eqnarray}
with the measurement strengths 
\begin{eqnarray}
\gamma_x = \int g_x (t) dt , \ x=a,b.   
\end{eqnarray} 
The measurements are assumed to be weak upon the fulfillment of a
certain weakness condition \cite{aharonov90}, which essentially
entails that the measurement strengths should be much smaller, in some
appropriate units, than the width of the corresponding (Gaussian)
pointer wave functions. It can be shown \cite{aharonov90} that it is 
only necessary to keep terms to first order in the measurement 
strengths, in this weak measurement limit.

After the pointers have interacted weakly with the qubits Alice
performs a post-selection of the state $\ket{\varphi_+}_{12}$ on her qubit
pair. Conditioned on the post-selection, one finds to first order in
$\gamma_a$ and $\gamma_b$, i.e., in the weak measurement limit, the
unnormalized state vector of the system and the pointers to be
\begin{eqnarray} 
\ket{\Gamma_{ps}} & \approx & \frac{1}{2} \ket{\varphi_+}_{12} 
\ket{\phi}_3 e^{-i\gamma_a A_w q_a} \ket{m_a} 
e^{-i\gamma_b B_w q_b}  \ket{m_b} 
\nonumber \\ 
 & & - \frac{i}{2} \ket{\varphi_+}_{12} \ket{\phi^{\perp}}_3  
\Big( \gamma_a \bra{\phi^{\perp}} A \ket{\phi} q_a \otimes I_b 
\nonumber \\ 
 & & + \gamma_b \bra{\phi^{\perp}} B \ket{\phi} I_a \otimes q_b \Big) 
\ket{m_a} \ket{m_b} ,  
\end{eqnarray}
where $X_w = \bra{\Psi} X |\Phi \rangle / \bra{\Psi} \Phi \rangle$,
$X=A,B$, are the weak values with $\ket{\Psi} = \ket{\varphi_+}_{12}
\ket{\phi}_3$, and $\ket{\phi^{\perp}}$ is orthogonal to $\ket{\phi}$. 
Using the explicit form of $\ket{\Phi}$ and $\ket{\Psi}$ we obtain
\begin{eqnarray} 
X_w = \bra{\phi} X \ket{\phi} . 
\end{eqnarray} 
By taking the partial trace over the qubits, the state $\rho_{ab}$ 
of the measuring devices becomes 
\begin{eqnarray}
\rho_{ab} & \approx & 
\frac{1}{4} e^{-i\gamma_a A_w q_a} \ket{m_a} \bra{m_a} 
e^{i\gamma_a A_w q_a} 
\nonumber \\ 
 & & \otimes e^{-i\gamma_b B_w q_b} \ket{m_b} 
\bra{m_b} e^{i\gamma_b B_w q_b} .  
\end{eqnarray} 
Thus, at the expense of entangled particle pairs, Alice and Bob can
measure expectation values with respect to the same unknown state
$\ket{\phi}$ as shifts in the pointer momentum, in the weak
measurement limit. This is identical to the result obtained if Alice
and Bob had performed their local weak measurements on the state
$\ket{\phi}\ket{\phi}$. In this sense, the above scheme simulates
perfect cloning of quantum states.

Note that Bob may perform his measurement as soon as he has received
his part in the entangled pair of qubits. He might even do this
measurement before Alice receives the state $\ket{\phi}$. After his
measurement Bob awaits the message from Alice which details whether 
the measurement result should be kept or not. By repeating this procedure
Bob can reconstruct his expectation value. This is in contrast with a
scheme based on teleportation. In such a scheme, Alice first makes a
weak measurement on the unknown state $\ket{\phi}$ followed by a
teleportation to Bob who makes his weak measurement on the teleported
state. In this way, Alice and Bob also obtain the weak values
$A_w=\bra{\phi} A \ket{\phi}$ and $B_w=\bra{\phi} B \ket{\phi}$, 
respectively. The difference is that the weak
measurements have to be sequential since Bob has to await the
arrival of the signal from Alice before he can do his measurement. In
the weak cloning scheme, however, the weak measurements are
`simultaneous' in the sense that there is no causal ordering between
the two measurement events.

The independence of the two measurement events can perhaps be
discerned more clearly in a slightly modified version of the weak
cloning scheme. As described above, Alice and Bob initially share an
entangled state over systems $2$ and $3$. Alice performs the
post-selection locally on her systems $1$ and $2$, and transmits the
result to Bob. One can consider an alternative with Bob in sole
possession of the entangled pair $2$ and $3$, whereas Alice only
possesses system $1$. As a consequence, the post-selection on system
$1$ and $2$ cannot be performed locally, as it directly involves both
Alice and Bob. In this alternative scheme, Alice and Bob do not even
share any correlation at the time of their weak measurements. 

By sharing another Bell state with a third party (Charlie), Bob can
post-select a Bell state conditioned on Alice's post-selection. By
performing a weak measurement of an observable $C$, say, Charlie can
reconstruct the weak value $C_w=\bra{\phi} C \ket{\phi}$ upon 
receiving the results of Bob's post-selection. Continuing in this 
way, the unknown state can be weakly cloned arbitrary many times, 
at the expense of an intensity loss that is exponential in the number 
of post-selection steps.

The weak cloning scheme can be seen as a parallelization of 
consecutive weak measurements. By a similar post-selection procedure
one can obtain parallelization of other operations, like other types
of measurements, or unitary transformations. It is important to note,
though, that the post-selection procedure results in that Bob operates
on the output of Alice, and similarly that Charlie operates on the
output of Bob. That Bob operates, or measures, on the output of Alice
clarifies the role of the weak measurement in the weak cloning
scheme. Since in the weak limit Alice's measurement does not perturb
the state, Bob can measure as if on the original input state. This
would not be the case if Alice made an ordinary type of measurement,
and similarly for Charlie if Bob made an ordinary measurement.
  
Generalization of the above scheme to systems described by $N\geq 3$ 
dimensional Hilbert spaces can be achieved by replacing the Bell 
state by a maximally entangled state of the type $\ket{\psi} = 
\sum_j \ket{j} \ket{j} /\sqrt{N}$. Alice then performs a 
post-selection by measuring an observable with $\ket{\psi}$ 
as a nondegenerate eigenvector. She communicates which of her systems 
pass the test to Bob, who in turn uses the corresponding data to 
obtain his weak value. 

Experimental realization of the weak cloning scheme is most likely to
be found for photonic systems. We note that two-photon systems have
already been used to demonstrate some properties of weak values
\cite{pryde04}. In view of the rapid development of techniques for
quantum control in general and for photons in particular, we believe
experimental test of weak cloning should be feasible in the near
future.

\end{document}